\documentclass[preprint,12pt]{aastex}
\usepackage{epstopdf,natbib}

\newcommand{\teff}{$T_{\rm eff}$}
\newcommand{\logg}{$\log{g}$}
\newcommand{\fsed}{$f_{\rm sed}$}

\newcommand{\name}{WISEP J061135.13$-$041024.0}
\newcommand{\namesh}{WISE 0611$-$0410}
\newcommand{\chisq}{$\chi^2$}
\newcommand{\kzz}{$K_{\rm zz}$}

\shorttitle{WISE 0611$-$0410AB}
\shortauthors{Gelino et al.}

\begin{document}

\title{\name AB: A $J$-Band Flux Reversal Binary at the L/T Transition}

\author{Christopher R.\ Gelino\altaffilmark{1,2}, R. L. Smart\altaffilmark{3}, Federico Marocco\altaffilmark{4}, 
J. Davy Kirkpatrick\altaffilmark{2}, Michael C. Cushing\altaffilmark{5}, Gregory Mace\altaffilmark{6}, Rene A. Mendez\altaffilmark{7},
C. G. Tinney\altaffilmark{8,9}, \and Hugh R. A. Jones\altaffilmark{4}
}

\altaffiltext{1}{NASA Exoplanet Science Institute, Mail Code 100-22, California Institute of Technology, 770 South Wilson Ave, Pasadena, CA 91125, USA}
\altaffiltext{2}{Infrared Processing and Analysis Center, Mail Code 100-22, California Institute of Technology, 1200 E. California Blvd., Pasadena, CA 91125, USA}
\altaffiltext{3}{Istituto Nazionale di Astrofisica, Osservatorio Astrofisico di Torino, Strada Osservatorio 20, 10025 Pino Torinese, Italy}
\altaffiltext{4}{Centre for Astrophysics Research, University of Hertfordshire, Hatfield AL10 9AB, UK}
\altaffiltext{5}{Department of Physics and Astronomy, MS 111, University of Toledo, 2801 W. Bancroft St., Toledo, OH 43606-3328, USA}
\altaffiltext{6}{Department of Physics and Astronomy, UCLA, 430 Portola Plaza, Box 951547, Los Angeles, CA 90095-1547, USA}
\altaffiltext{7}{Universidad de Chile, Departamento de Astronom{\'i}a, Casilla 36-D, Santiago, Chile}
\altaffiltext{8}{School of Physics, University of New South Wales, Sydney, NSW 2052, Australia}
\altaffiltext{9}{Australian Centre for Astrobiology, University of New South Wales, Sydney, NSW 2052, Australia}

\begin{abstract}
We present Keck II laser guide star adaptive optics observations of the brown dwarf \name\ showing it is a binary with a component separation of 0\farcs4.  This system is one of the six known resolved binaries in which the magnitude differences between the components show a reversal in sign between the $Y$/$J$ band and the $H$/$K$ bands.  Deconvolution of the composite spectrum results in a best fit binary solution with L9 and T1.5 components.  We also present a preliminary parallax placing the system at a distance of 21.2$\pm$1.3~pc.  Using the distance and resolved magnitudes we are able to place \name AB on a color-absolute magnitude diagram, showing that this system contributes to the well-known ``$J$-band bump'' and the components' properties appear similar to other late-type L and early-type T dwarfs.   Fitting our data to a set of cloudy atmosphere models suggests the system has an age $>$1 Gyr with \namesh A having an effective temperature (\teff) of 1275-1325 K and mass of 64--65 $M_{\rm Jup}$, and \namesh B having \teff = 1075--1115 K and mass 40--65 $M_{\rm Jup}$.
\end{abstract}


\section{Introduction}
The $H$-band flux of stellar objects changes by nearly 12 orders of magnitude from O stars down to the Y dwarfs \citep{2012ApJ...753..156K}.  The progression of absolute $H$ magnitude ($M_H$) along this sequence is fairly smooth except for two major kinks\footnote{There could be a third kink in this sequence in the Y dwarfs \citep{2012ApJ...753..156K}}.  The first occurs in the early-type M dwarfs, where the formation of H$_2$ at these temperatures results in a short plateau in $M_H$ \citep{1976A&A....48..443M,1976ApJ...208..399M}.  
The other kink lies at the transition between the L and T dwarfs.  It is here that the absolute $H$ magnitude shows a distinct flattening before turning downward at the early-mid type T dwarfs \citep{2003AJ....126..975T}.  The source of this flattening, and the slight increase in absolute $J$ at these same spectral types, is not understood, but could be related to clouds, as discussed below.

Clouds are required in the atmospheric models of L dwarfs in order to replicate the observed spectral and photometric trends.  Conversely, models to match the observed properties of T dwarfs require the clouds to be completely dissipated, leaving the atmospheres clear of condensates.  The transition from the cloudy L dwarfs to the clear T dwarfs is not well understood and has been notoriously difficult to model.  A gradual lowering of a cloud deck through the L to T cooling sequence using a single sedimentation efficiency \citep[\fsed;][]{2001ApJ...556..872A} produces a shift in the near-IR magnitudes that is too slow to account for the observed colors and absolute magnitudes of L and T dwarfs \citep{2004AJ....127.3553K}.  However, if \fsed\ increases from $\sim$2 for the L dwarfs to infinity for the T dwarfs, giving the T dwarfs cloud-free atmospheres, then the bump in absolute $J$ magnitude can be modeled.  Conceptually, the change in \fsed\ means that condensates in the L dwarfs would precipitate out of the clouds like a light rain shower.  As the object cools to later spectral types, the clouds fall deeper in the atmosphere and the condensates rain out progressively harder and harder, until the clouds drop below the photosphere and the atmosphere appears cloud-free.  

The rapid clearing of the clouds and transition from L dwarf to T dwarf can also be explained by holes in the clouds \citep{2001ApJ...556..872A,2002ApJ...571L.151B,2010ApJ...723L.117M,2013arXiv1311.2085S}.  Small holes in the cloud deck allow hotter flux from deeper layers to emerge, analogous to Jupiter's 5 $\micron$ ``hot-spots'' \citep{1974ApJ...188L.111W}.  The presence of the holes results in a slightly brighter luminosity and bluer $J-K$ color.  The coverage area of the holes would rapidly progress from 0\% for the cloudy L dwarfs to 100\% for the clear T dwarfs.   

In addition to the bump in absolute $J$ magnitude, the L-T transition region is also characterized by a resurgence in the strength of FeH absorption at 9896~\AA\ \citep{2002ApJ...571L.151B}.  This feature becomes weaker through the L dwarf sequence, nearly disappearing in L8 dwarfs, but then reappears in the mid-T dwarfs.  Both the ``sudden downpour'' and the ``cloud disruption'' models can account for this resurgence in the FeH feature.  In the sudden downpour model, the increase in \fsed\ means that the Fe cloud becomes thinner and drops deeper into the atmosphere.   Consequently, the amount of FeH that is visible increases, thereby increasing the strength of the absorption feature \citep{2008ApJ...678.1372C}.  Conversely, the holes in the cloud disruption model allows gaseous FeH below the cloud deck to become visible \citep{2002ApJ...571L.151B}, which results in stronger FeH features in the observed spectra of early-T dwarfs.  

Although some objects on the bright end of the bump may be unresolved binaries \citep{2006ApJ...640.1063B}, these ``crypto-binaries'' cannot account for all objects in the bump.  Several binary systems, typically consisting of a late-L to early-T dwarf primary and an early- to mid-T dwarf secondary, have been discovered in which the secondary component is brighter than the primary in the 1.0-1.3~$\micron$ region (see Table~\ref{tab_jflip}).    These systems indicate that the brightening in $J$-band from late-type L to early-type T is an intrinsic property of these objects and not the result of some selection effect.  For example, the theory that the apparent brightening in the color-magnitude diagram is caused by objects with widely varying gravities \citep{2003ApJ...585L.151T} can be discounted because such a scenario requires objects at vastly different ages, which is not likely in these presumably coeval binary systems.

In this paper we present the sixth binary showing a flux reversal in the 1.0-1.3~$\micron$ region.  \name\ (\namesh, hereafter) was first presented in \citet{2011ApJS..197...19K} as a T0 with a spectro-photometric distance of 17.7~pc.  It was discovered in a Wide-field Infrared Survey Explorer \citep[WISE;][]{wright2010} search for bright sources  with brown dwarf-like colors and no counterpart in the Two Micron All Sky Survey catalog \citep[2MASS;][]{2006AJ....131.1163S}.  Such properties are indicative of a source that either has a large proper motion or one with a very red $J-W2$ color.  In the case of \namesh\ the source {\it is} present in the 2MASS catalog (See Table~\ref{tab_phot} for a summary of the photometry of this object), but its position has moved by more than 3$\arcsec$ between the two surveys.  Follow-up observations discussed here have accurately measured the proper motion and provide a first determination of the parallax.  We also show resolved photometry and spectral deconvolution that clearly demonstrate the flux reversal in the 1.0-1.3~\micron\ region.

\section{Observations\label{obs}}

\subsection{NTT}
Parallax observations were carried out using the European Southern Observatory (ESO)
3.5m New Technology Telescope (NTT) and its infrared
spectrograph and imaging camera SofI
\citep[Son of ISAAC;][]{1998Msngr..91....9M} as part of program 186.C-0756:
{\it NTT Parallaxes of Southern Extremely Cool objects.}
The observing, analysis, and reduction procedures are
described in detail in \citet{2013MNRAS.tmp.1584S}
and are only briefly summarized here. The
observations were obtained in SofI's large field mode,
providing a 4$\farcm$9$\times$4$\farcm$9 field with
0\farcs288 pix$^{-1}$ in the $J$ band.  Each epoch consists of 18
dithered observations of 4$\times$30s
at each pointing for a total exposure time of 36 minutes. The
observations were combined using the {\it jitter} routine
of the {\it Eclipse} \citep[][version 5.0]{1997Msngr..87...19D} package.  Objects in the field were found and
centroided using the Cambridge Astrometry Survey UnitÕs
{\it imcore} maximum likelihood barycenter software
(CASUTOOLS\footnote{\url{http://apm49.ast.cam.ac.uk/surveys-projects/software-release}}, v 1.0.21).

During nights of exceptional seeing, one of us (RAM), noted
that the point spread function (PSF) of  \namesh\ was slightly elliptical compared to other objects in the field.
This prompted the high-resolution follow-up and spectral analysis discussed here. 

\subsection{Keck II/NIRC2}
High resolution images of \namesh\ were obtained using the Keck II laser guide star adaptive optics (LGS-AO) system \citep{wizinowich2006,vandam2006} with NIRC2 on the nights of 2012 January 13, 2012 April 15, 2012 September 06, 2012 November 29, and 2013 September 21 (see Table~\ref{tab_obs}).  Our target was too faint in $R$-band to serve as a tip-tilt reference star for the wavefront corrector, so we used the USNO-B star 0858-0074221 \citep{monet2003} with $R$=16.5 and located 35$\arcsec$ from our target.  Our interest was focused on the target and its immediate vicinity, so we used the narrow camera mode with a nominal pixel scale of 10 mas pix$^{-1}$ and single-image field of view of 10$\arcsec$.  We used a 3-point dither pattern to avoid the high noise levels in the lower-left quadrant of the array.  This pattern was repeated with different offsets to build up a longer exposure.  The Mauna Kea Observatories \citep[MKO;][]{tokunaga2002} $Y$, $J$, $H$, $K$, and $K_s$ filters were used on one or more nights (Table~\ref{tab_obs}).

The images were reduced using scripts in the IDL (Interactive Data Language) environment.  A dark frame was first subtracted from each science frame.  Then, a sky frame, created from the median average of all images acquired for \namesh\ exclusive of the frame being reduced, was subtracted.  The sky-subtracted frames were then divided by a dome flat.  Finally, the images were shifted to align the target to a common location and a deep mosaic was created from the median average of the stack.  For the single image obtained on 2012 November 29, we could only dark subtract and flat-field the science frame.  Fortunately, since our interest in that data was to follow the relative motions of the objects, the lack of a sky frame did not impede our goals.  Table~\ref{tab_obs} lists the total exposure times for each of the mosaics, as well as the full-width at half-maximum (FWHM) and Strehl ratios for the objects in frames.

\begin{figure}
\plotone{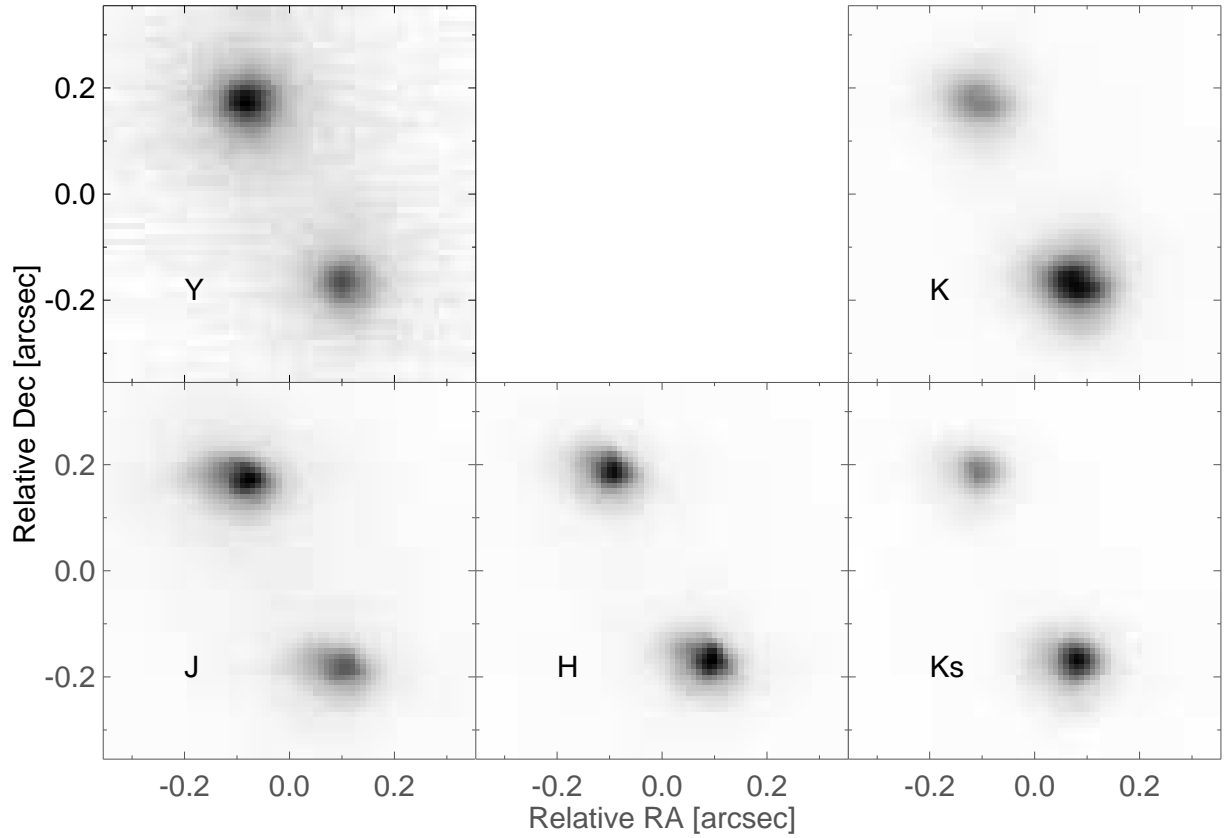}
\caption{Keck LGS-AO NIRC2 images of \namesh AB obtained on 2012 April 15 (bottom row) and on 2013 September 21 (top row).  The images are shown with North up and East to the left.\label{fig_mosaic}}
\end{figure}

The $Y$, $J$, $H$, $K$, and $K_s$ mosaics from the 2012 April 15 and 2013 September 21 nights are shown in Figure~\ref{fig_mosaic}.  As was suspected from the NTT data, this source is easily resolved into two components.  Visual inspection of the images shows that the relative brightness changes sign between the $Y$ \& $J$ and $H$, $K$, \& $K_s$ images.  We discuss the implications of this below.

\section{Analysis}
\subsection{Parallax and Proper Motion}
\begin{figure}
\epsscale{1.0}
\plotone{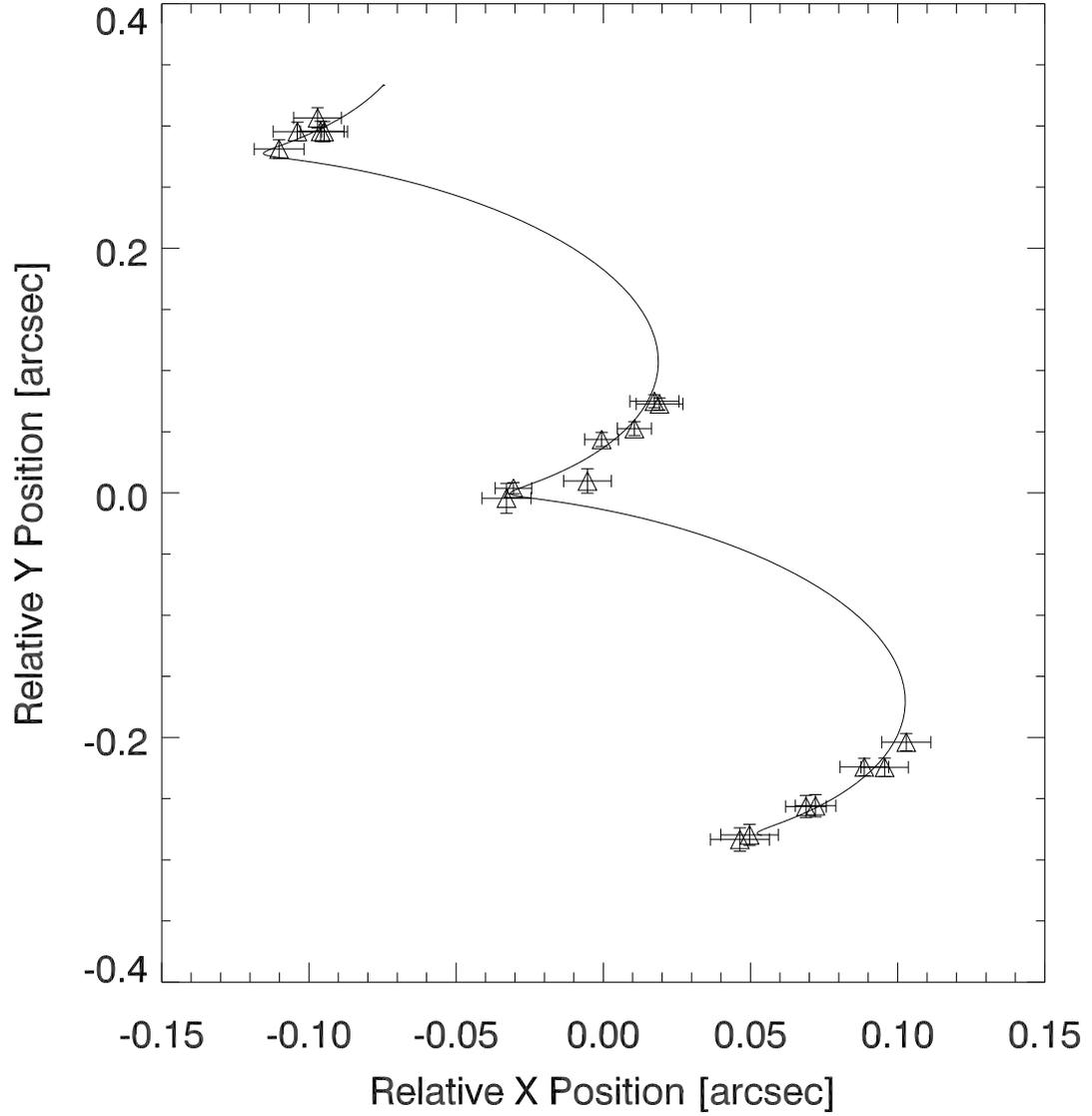}
\caption{The apparent motion of \namesh AB on the sky over the nineteen NTT epochs.  The offsets are relative to the middle observation.  Also drawn is the best-fit curve to the parallax and proper motion.  \label{fig_plx}}
\end{figure}

We have 20 observations of \namesh\ from
2010 December 23 through 2013 February 8. An observation in the
middle is chosen as the master frame and the measured
coordinates transformed to a standard system using 
2MASS objects in the field. All other observations are
transformed to this standard system using all common objects
and a simple linear transformation. Once all frames are on
the same system, the motion of the target is fit for the
astrometric parameters: position, relative proper motion,
and relative parallax.  
This procedure is iterated while removing reference stars, which have brightnesses straddling that of \namesh, with high scatter about the transformation. One epoch with an anomalously large residual ($\sim$3$\sigma$) compared to the best fit was also dropped.
After this iteration a correction of the
relative parallax to an absolute value is calculated
using the galaxy model of \citet{men96}. The resulting astrometric motion of \namesh\ across the sky from these 19 epochs, along with
our fit, is shown in Figure~\ref{fig_plx} and
the astrometric parameters are produced in Table~\ref{tbl_plx}.

\subsection{Resolved Photometry\label{sec-res_phot}}
Our Keck imaging easily resolves \namesh\ into its component sources, allowing for simple calculation of the magnitude differences in the five filters.  To derive the MKO-based photometry of the two components, we need MKO magnitudes for the composite source in those filters.  MKO $J$, $H$, and $K$ images of \namesh\ were obtained by the UKIRT Infrared Deep Sky Survey \citep[UKIDSS;][]{2007MNRAS.379.1599L} Galactic Clusters Survey.  The $H$ and $K$ data were obtained from the public Data Release 9; $J$-band data comes from unreleased images observed on 2012 Dec 09.  These images show an elliptical source, similar to what is seen in the NTT data.  No MKO $Y$ or $K_s$ images of this field have been obtained.  This UKIDSS photometry, as well as the 2MASS photometry, is given in Table~\ref{tab_phot}.

The separation between the two components is large enough that basic aperture photometry can be used to measure the magnitude differences.  We use the data from 2012 April 15 and 2013 September 21 for the photometry since they are of the highest quality and include the largest set of filters.  The photometry is performed on the mosaicked images (Figure~\ref{fig_mosaic}) and is presented in Table~\ref{tab_binprop}.  We denote the source that is fainter at $Y$ and $J$ as component A in this system.

\subsection{Resolved Astrometry and Companionship}
We have observed \namesh AB using high-resolution imaging on five different epochs with the purposes of confirming a common proper motion between the sources and searching for any orbital motion.  For a given epoch, we measure the separation in RA and Dec from source A to source B on all of the individual images.  We then take the average separations from the measurements and assign the standard deviation as the error in those quantities.  Since the 2012 September 06 epoch only contained a single, good quality image, we conservatively set the separation error in each axis to be 10 mas ($\approx$1 pixel).  

\begin{figure}
\epsscale{1.0}
\plotone{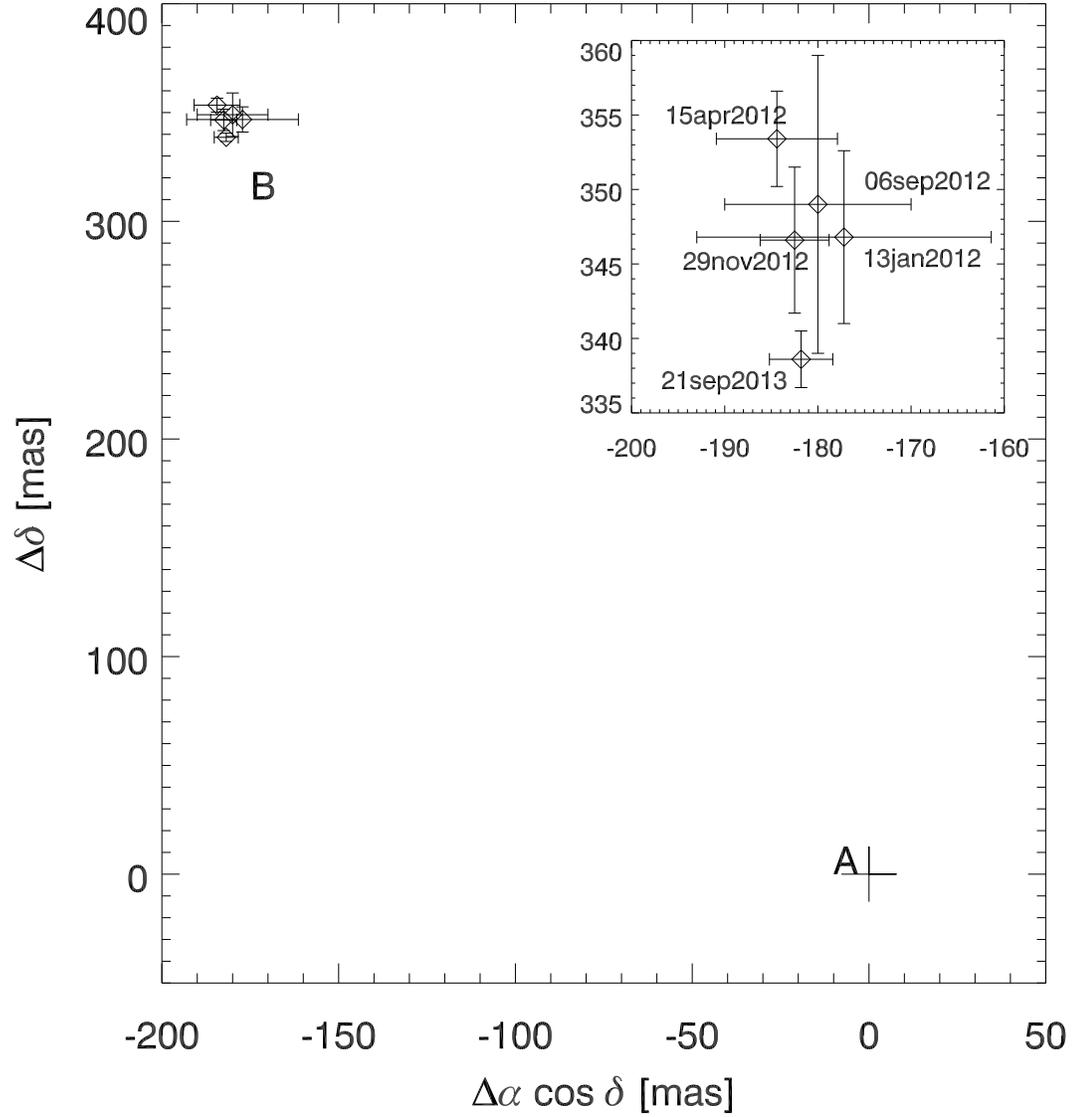}
\caption{Motion of \namesh B with respect to \namesh A from the NIRC2 data.  The inset shows the five epochs in more detail.  We do not detect any significant motion of the companion, relative to component A, throughout our NIRC2 observations. \label{fig_orbit}}
\end{figure}

The final measurements are shown in Table~\ref{tab_sep} and graphically in Figure~\ref{fig_orbit}.    If source B was a background object, then its position would have changed by 480 mas over the course of the NIRC2 observations.  A positional offset that large is easily detectable in these high quality observations.  However,  there is no significant change in the relative position of source B over the five epochs.  Therefore, we can confidently confirm that these sources are physically bound and, additionally, no orbital motion is yet detected.  The parallax and proper motion fit also do not show any evidence for binary motion over more than 2 years of NTT observations.

\subsection{Composite Spectrum Deconvolution}
To determine the spectral types of the individual components of \namesh AB we used a three-step deconvolution procedure, similar to that described in \citet{2013MNRAS.430.1171D}. First we fit the entire spectrum (0.85-2.45 $\mu$m) of our target \citep{2011ApJS..197...19K} with the near-infrared spectroscopic standards defined in \citet{2010ApJS..190..100K} and \citet{2006ApJ...637.1067B} to determine the best-fit standard via $\chi^2$ minimization. The best-fit standard selected this way is SDSS~J120747.17+024424.8 with spectral type T0, identical to the spectral type found by \citet{2011ApJS..197...19K}.

\begin{figure}
\epsscale{1.0}
\plotone{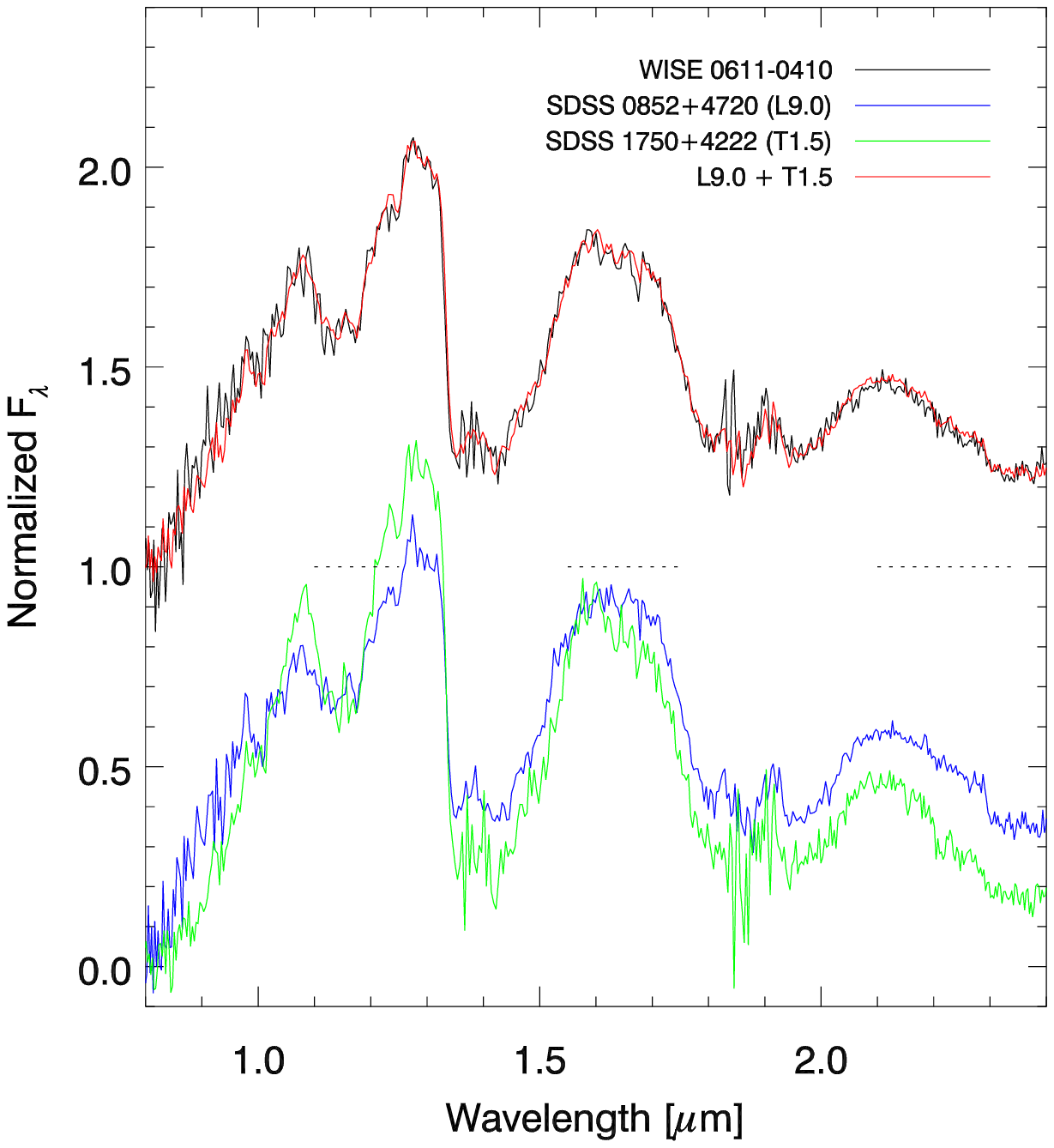}
\caption{Near-IR spectrum of \namesh AB from \citet[black line]{2011ApJS..197...19K}.  Also shown are the spectra of the template L9 dwarf SDSS J085234.90+472035.0 \citep[blue line;][]{2010ApJ...710.1142B} and the T1.5 dwarf SDSS J175024.01+422237.8 \citep[green line;][]{2010ApJ...710.1142B}, whose convolution (red line) provides the best fit to the observed \namesh AB spectrum.  The spectral regions used in the binary fitting are denoted by the dotted lines.  The flux reversal between 1.0 and 1.3 $\micron$ observed in the NIRC2 photometry is also evident in the comparison of the template spectra.\label{fig_spec}}
\end{figure}

Then we repeated the fit using a set of synthetic binaries. Synthetic binaries were created combining the template spectra taken from the SpeX-Prism library\footnote{\url{http://pono.ucsd.edu/$\sim$adam/browndwarfs/spexprism/}}, scaling them to the appropriate relative flux level using the spectral type$-$absolute magnitude calibration presented in \citet{marocco2010}. We selected only those binaries giving a lower $\chi^2$ compared to the standard template. Among those binary pairs, the best-fit binary was found by only fitting three regions of the spectrum: 1.10-1.25 $\mu$m (containing CH$_4$ and H$_2$O absorption bands), 1.55-1.75 $\mu$m (containing the CH$_4$ absorption band), and 2.10-2.35 $\mu$m (containing the CO absorption band). The features within these intervals change significantly at the transition between L and T types, and are therefore the most suitable to identify and deconvolve unresolved L/T pairs. The best-fit template identified with this procedure (see Figure~\ref{fig_spec}) consists of a L9 dwarf (SDSS~J085234.90+472035.0) and a T1.5 dwarf (SDSS~J175024.01+422237.8).

We tested the significance of our deconvolution using an F-test to compare the result of the fit with synthetic binaries against the fit with the standard template alone \citep{2010ApJ...710.1142B}.  We use an F distribution threshold of 99.5$\%$, which translates to a critical value ($\eta_{crit}$) of 1.41. If the ratio of the two $\chi^2$ (defined as $\eta$) is larger than $\eta_{crit}$, then the combined template provides a better fit at the 99.5$\%$ confidence level.  We obtained $\eta$ = 1.73 for the binary fit using L9 and T1.5 templates and, therefore, assign those spectral types to the two components.

\subsection{Discussion}
\subsubsection{Comparison to Cloudy Models}
A true test to firmly pin down the atmospheric properties of these brown dwarfs requires high resolution, high signal-to-noise ratio spectroscopy.  With such data it would be possible to distinguish the effects of gravity and non-equilibrium chemistry that are discernible in the models.  With no resolved spectroscopy we will attempt to estimate the physical properties of \namesh AB using our measured photometry and distance.

We use the atmosphere models of \citet{2008ApJ...689.1327S} and \citet{2009ApJ...702..154S} for comparison to our binary.  These models  incorporate a variety of cloud sedimentation efficiencies \citep[\fsed;][]{2001ApJ...556..872A} that dictate the size of the condensate particles and the thickness of the cloud deck.  Small values of \fsed\ produce thick clouds comprised of small particles, whereas large values of \fsed\ produce thin clouds with large particles.  These models also include a vertical mixing component through the eddy diffusion coefficient, \kzz.  This vertical mixing can throw the observable abundances of molecules such as CO, CH$_4$, H$_2$O, NH$_3$, and N$_2$ out of chemical equilibrium by dredging up molecules favored at high temperatures and pressures (eg., N$_2$ and CO) and mixing them higher in the atmosphere, where cooler temperature molecules (eg., CH$_4$, H$_2$O, NH$_3$) are favored.  If the replenishment of the ``hot'' molecules is faster than the reaction to convert them to the ``cold'' molecules, then there will be an over-abundance of CO compared to CH$_4$, for example.  Values of \kzz\ are typically $10^4$-$10^6$ cm$^2$ s$^{-1}$; in our model suite $\log$ \kzz\ is ``0'' (representing the chemical equilibrium case and {\it not} \kzz=1), 2, 4, and 6.  

It is important to note that the models we are utilizing are atmosphere models computed at a fixed grid of \logg\ and \teff.  The evolutionary models of \citet{2008ApJ...689.1327S} are then used to compute the mass. These evolutionary models are computed for clear atmospheres (not used in our analysis) and cloudy models with \fsed=2.  The masses and ages for other values of \fsed\ are approximate and based on \fsed=2 models with similar \teff\ and \logg.  Furthermore, although we show models with \logg\ as high as 5.477 (cm s$^{-2}$), these models produce objects that are physically impossible \citep{2008ApJ...689.1327S}.  In fact, the maximum value of \logg\ for a cloudy set of models is 5.38 (cm s$^{-2}$).

\begin{figure}
\epsscale{1.0}
\plotone{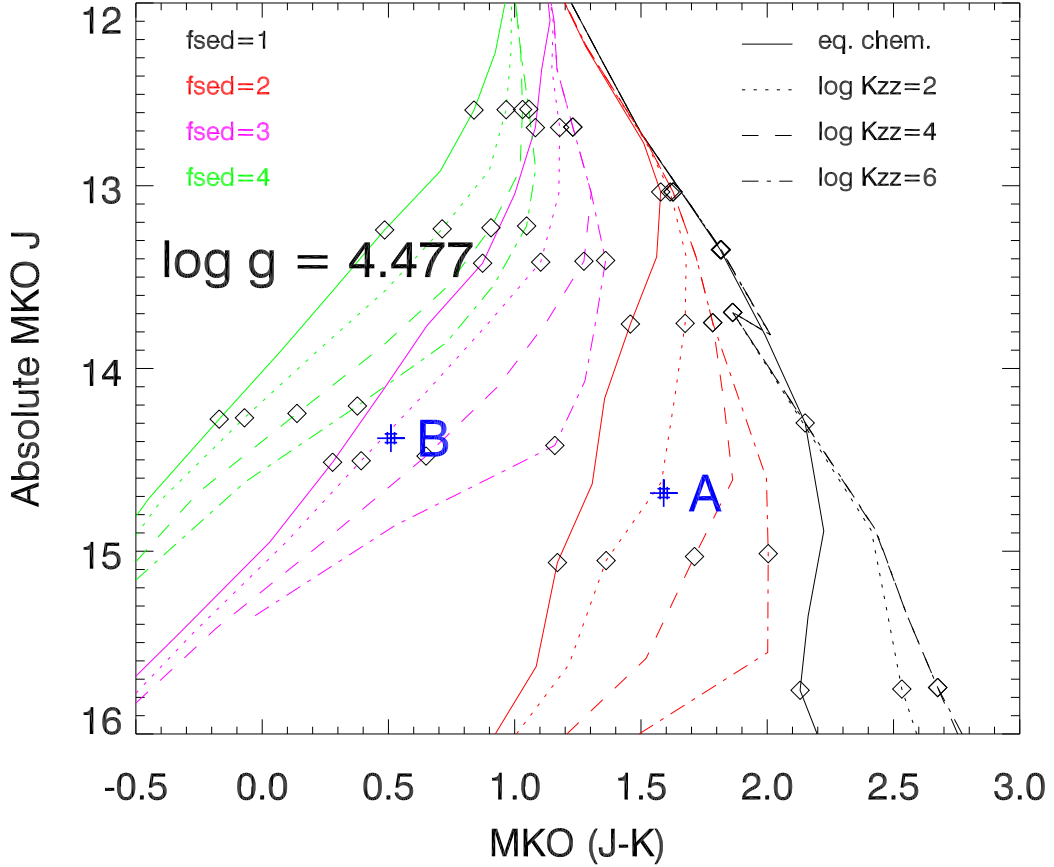}
\caption{Absolute MKO $J$ magnitude as a function of MKO $J-K$ for a suite of models with \logg\ = 4.477 (cm s$^{-2}$).  The atmosphere models include \fsed=1,2,3,4 with equilibrium chemistry and 3 levels of non-equilibrium chemistry mixing ($\log$ \kzz=2,4,6).  Effective temperatures vary from model to model but generally range from 500-2400K for the equilibrium chemistry models and 900-1800K for the $\log$ \kzz=6 models.  For reference, the locations of the \teff =1500, 1300, and 1000 K models are plotted with diamonds (top to bottom).
\label{fig_grav45_model}}
\end{figure}
\begin{figure}
\epsscale{1.0}
\plotone{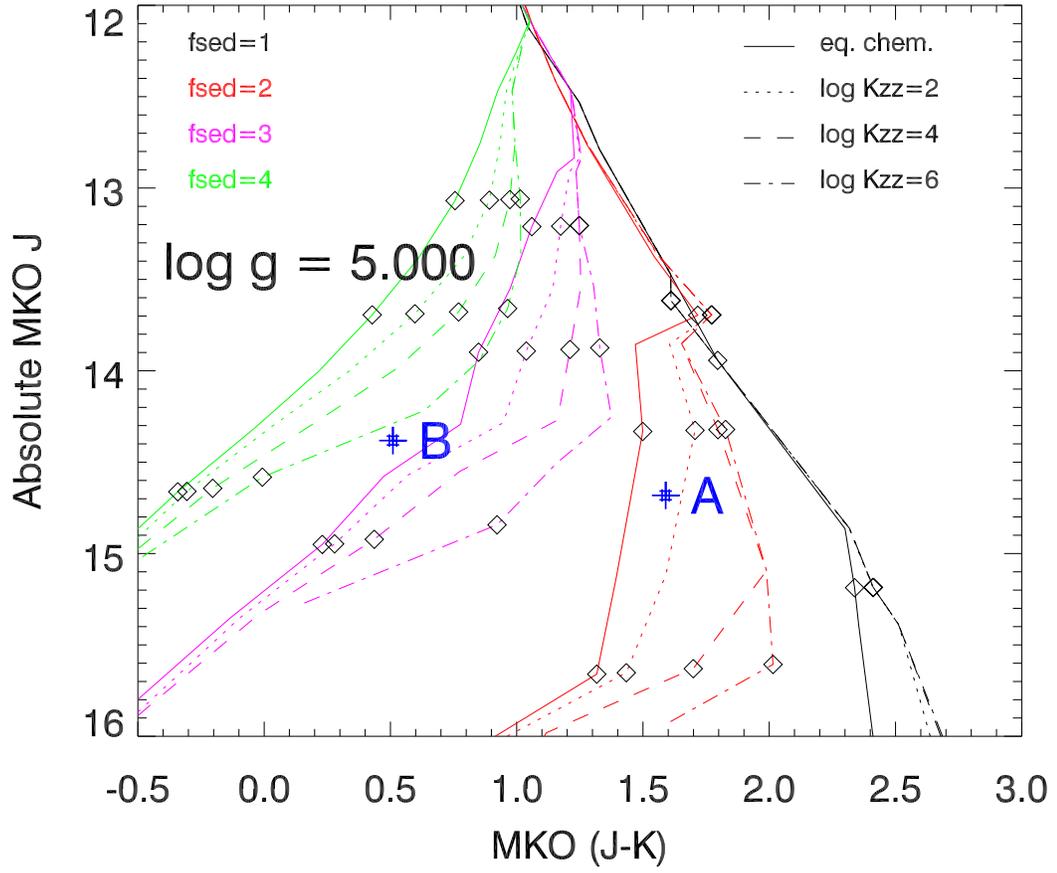}
\caption{Same as Figure~\ref{fig_grav45_model} except for \logg\ = 5 (cm s$^{-2}$). 
\label{fig_grav5_model}}
\end{figure}
\begin{figure}
\epsscale{1.0}
\plotone{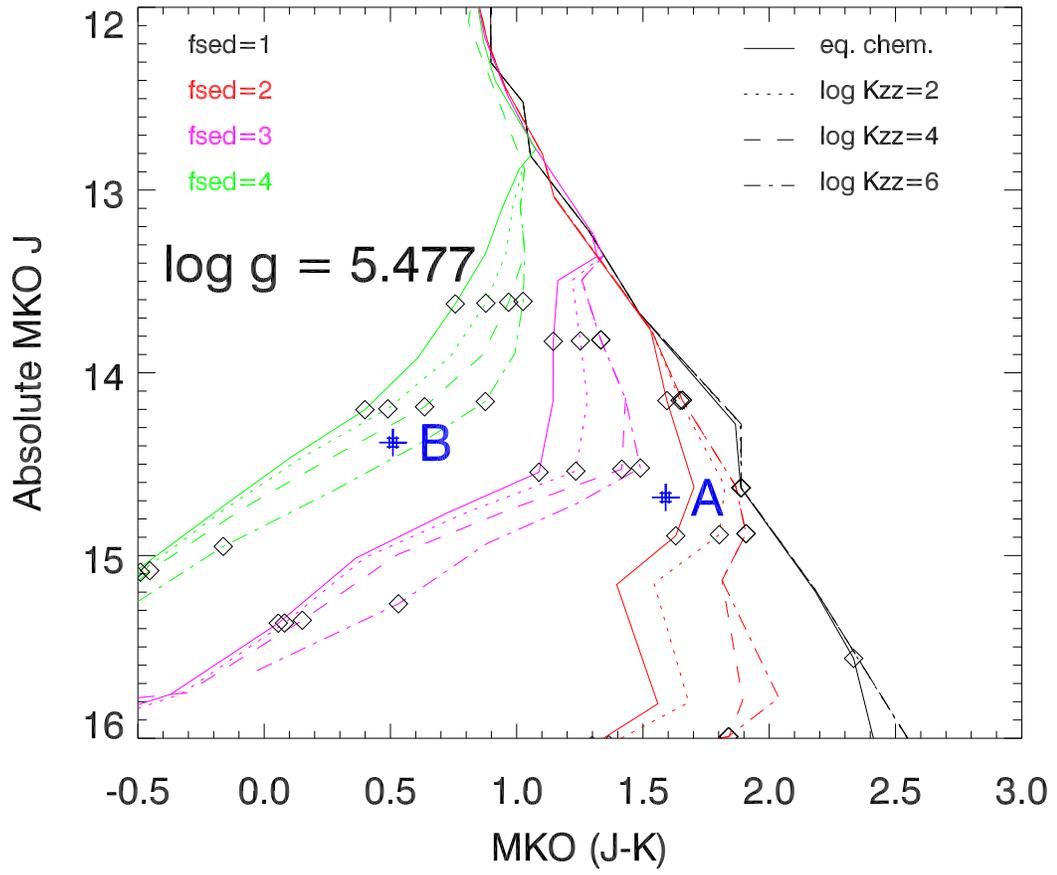}
\caption{Same as Figure~\ref{fig_grav45_model} except for \logg\ = 5.477 (cm s$^{-2}$).
\label{fig_grav55_model}}
\end{figure}

Figures~\ref{fig_grav45_model}, \ref{fig_grav5_model}, and \ref{fig_grav55_model} show $M_J$ as a function of $J-K$ for models at a variety of effective temperatures, $\log$ \kzz, and \fsed\ with \logg=4.477, 5.0, and 5.477 (cm s$^{-2}$).  As shown in these figures, increasing the model gravity tends to shift absolute magnitudes to fainter values and the $J-K$ color to the blue.  Likewise, models with a higher sedimentation efficiency are bluer in $J-K$ than lower \fsed\ models since the high \fsed\ models more closely resemble cloud-free atmospheres.   Finally, increasing \kzz\ tends to produce redder $J-K$ colors while leaving $M_J$ relatively unchanged.  

Since the grid of \teff\ and gravities can be somewhat coarse for a given value of \fsed\ and $\log$ \kzz\ we have constructed a 2-d interpolation using steps of 10 K and 0.1 (cm s$^{-2}$) for the \teff\ and \logg\ axes, respectively.  We acknowledge that this interpolated grid of models may not reproduce a model computed at each \teff\ and $\log g$, but feel that the interpolation does a sufficient job of estimating the actual model properties.  

To find the best fitting model, we minimize $\chi^2$ as given in the following equation   

\begin{equation}
\chi^2 = \sum\limits_{i}^{J,H,K} \frac{(M_{i,obs} - M_{i,model})^2}{\sigma^2 (M_{i,obs})} ,
\end{equation}

\noindent where $M_{i,obs}$ and $\sigma^2 (M_{i,obs}$) are the absolute magnitude and the standard deviation of the observed absolute magnitude, respectively, and $M_{i,model}$ is the absolute magnitude from the model.  The $\chi^2$ minimization is performed for each discrete set of \fsed\ and \kzz\ models.  Errors in the best fit parameters (mass, \teff, and \logg) are computed by randomly and independently varying the $J$, $H$, and $K$ absolute magnitudes over the range expected from their errors.

\begin{figure}
\epsscale{1.0}
\plotone{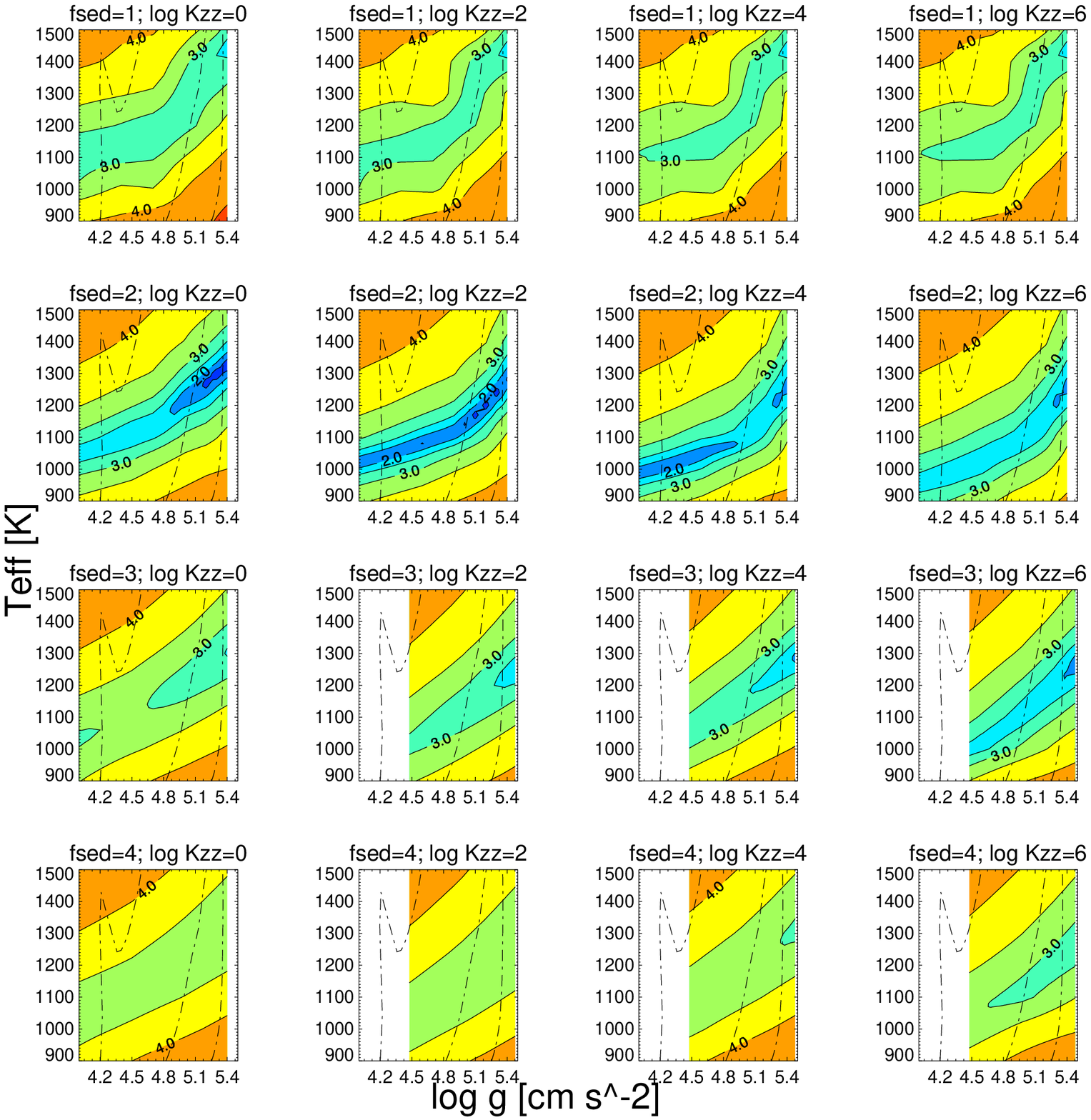}
\caption{Contours of model fitting $\log$ \chisq\ as a function of \teff\ and \logg\ for the 16 combinations of \fsed\ and $\log$ \kzz\ for \namesh A.  The dot-dashed lines denote \fsed=2, $\log$ \kzz=0 objects with ages 0.1, 1, and 6 Gyr (left to right).  The upturn in \teff\ from $\approx$1250 K to $\approx$1450 K for the 0.1 Gyr model is caused by deuterium fusion in the cores of these objects.  The contour levels and associated colors are the same in all panels.  The $\log$ \kzz=0 model represents the equilibrium chemistry case.  Note that the gravities for the \fsed=3 \& 4, $\log$ \kzz=6 do not span as wide of a range as the other model cases.
\label{fig_contour_a}}
\end{figure}

\begin{figure}
\epsscale{1.0}
\plotone{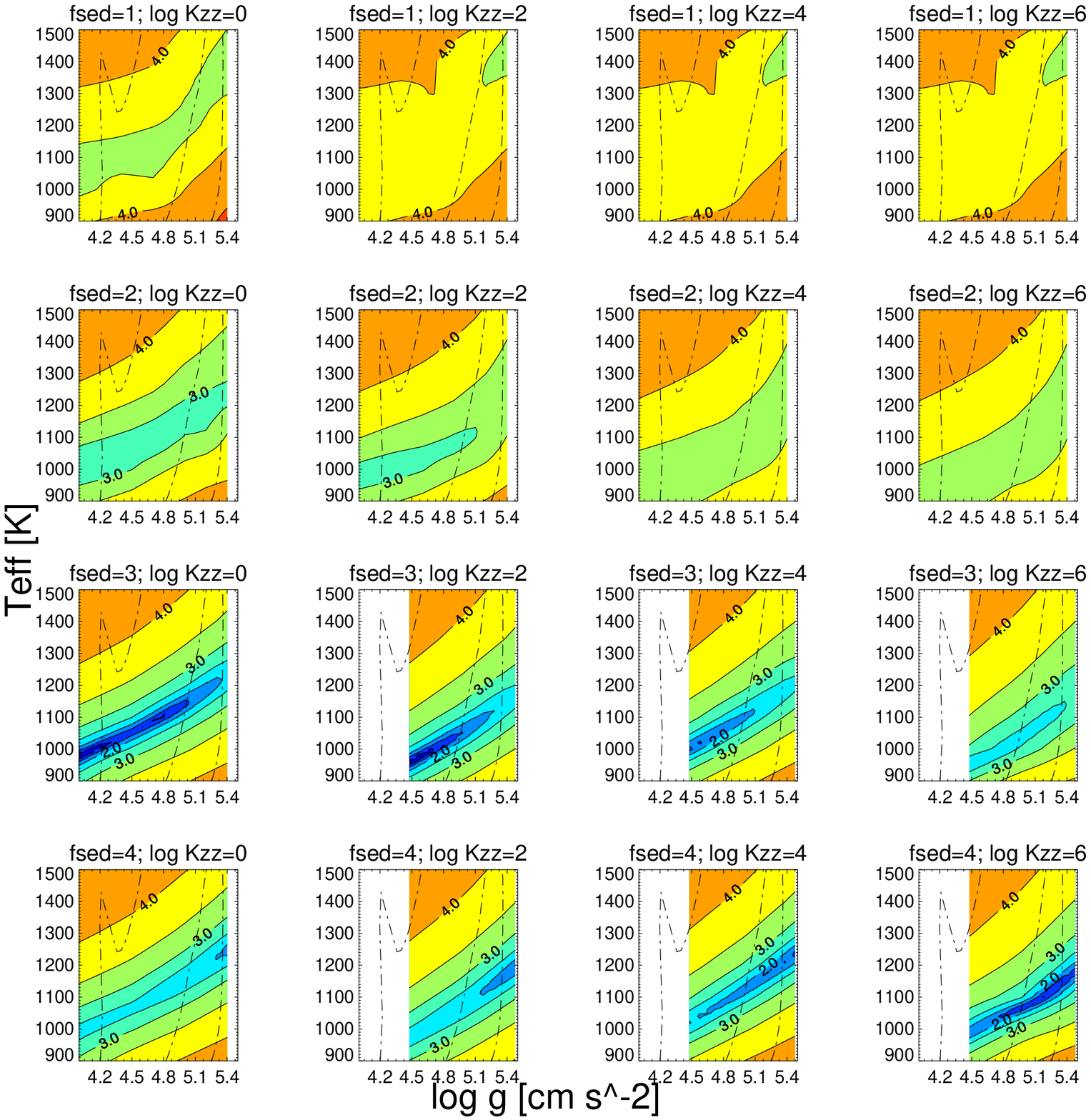}
\caption{Same as Figure~\ref{fig_contour_a} except for \namesh B.
\label{fig_contour_b}}
\end{figure}

The results of the \chisq\ minimization calculations as a function of \teff\ and \logg\ are shown in Figures~\ref{fig_contour_a} \&~\ref{fig_contour_b} for each set of \kzz\ and \fsed\ model.  Also shown on these plots are the loci of \fsed=2 models having ages of 0.1, 1, and 6 Gyr.  In general, \namesh A (Figure~\ref{fig_contour_a}) is fit well by \fsed=2 models having relatively high gravity (\logg$\gtrsim$5 [cm s$^{-2}$]) and \teff $\approx$1200-1300 K.  The corresponding mass for these models is 50-65 $M_{Jup}$  with ages $>$1 Gyr.

Conversely, the ranges of \logg\ and \teff\ of the best fitting models for \namesh B greatly depend on the cloud and mixing properties (Figure~\ref{fig_contour_b}).  Strictly speaking, the best-fitting models are with \fsed=3 and very low gravity.  The ages inferred from these models ($<$1 Gyr) are inconsistent with the ages for the \namesh A models.  The \fsed=4 models that provide a good fit to \namesh B have larger gravities and older ages, but do not fit the data quite as well as the \fsed=3 models.  To determine which set of models are a better physical fit to \namesh AB, we need to look at other observables that constrain the age of the system.

The space motion of \namesh AB effectively rules out membership of any known young moving group.  Using the method from \citet{marocco2010}, and assuming a radial velocity range of $\pm$100 km s$^{-1}$, we find that \namesh AB has a low probability of $\sim$~12\% of being younger than 500 Myr and $\sim$7\% of being younger than 100 Myr.  The composite spectrum does not show any indicators that this system is very young, so we can assume that the age of the system is more than a few hundred million years old \citep{kirk2005}.  This effectively rules out the low-gravity, \fsed=3 models for \namesh B, so we adopt the higher gravity, older \fsed=4 models.


The parameters for the best-fitting models with age$>$1 Gyr are shown in Table~\ref{tab_best_fit_models}.   \namesh A is best fit by a model with \teff$\approx$1300 K, whereas \namesh B is best fit by a model with \teff$\approx$1100 K.  The effective temperatures for both objects are largely consistent with those expected for late-type L dwarfs and early-type T dwarfs \citep{2009ApJ...702..154S} .  Likewise, the cloud properties are also not unusual for these types.  Being a late-type L dwarf, \namesh A is expected to have thick clouds covering most of its atmosphere.  Models with \fsed=2 are generally the best types for fitting late-type L dwarfs \citep{2008ApJ...689.1327S}.  Early-mid T dwarfs require models that are less cloudy than the L dwarfs, making an \fsed=4 model appropriate.  However, an alternate model can also be used to characterize \namesh B (Section~\ref{sec:partly_cloudy}).  

Finally, one interesting aspect of the model fitting is the very different levels of atmospheric mixing implied for the two objects.  While this could represent a genuine physical difference between these objects, it is impossible to make this assessment based entirely on the broad-band photometry in three filters.  High resolution spectroscopy of both components is needed to truly ascertain if either object shows non-equilibrium chemistry.

\subsubsection{Comparison to Partly Cloudy Models \label{sec:partly_cloudy}}
\begin{figure}
\epsscale{1.0}
\plotone{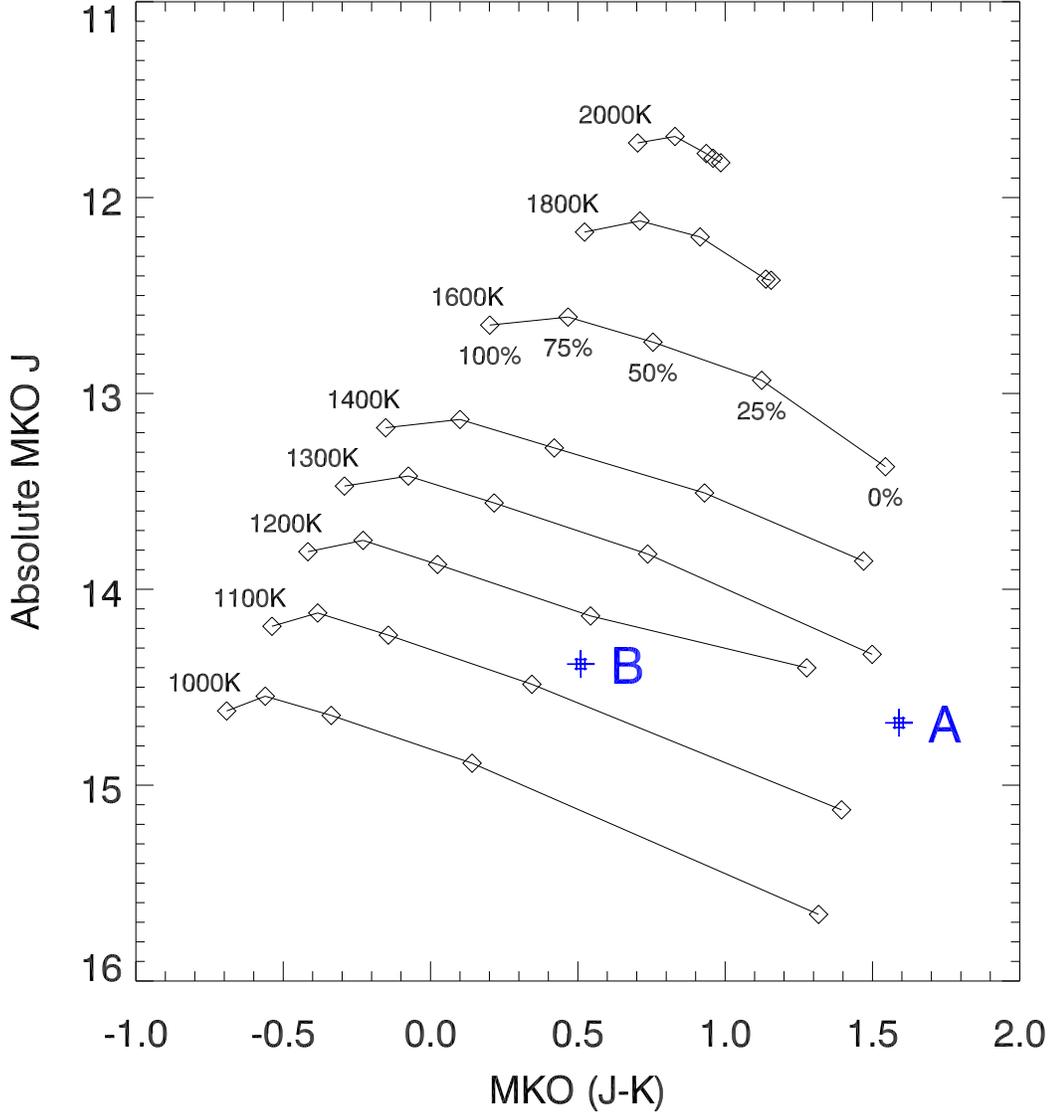}
\caption{Absolute MKO $J$ magnitude as a function of MKO $J-K$ for the partly cloudy models of \citet{2010ApJ...723L.117M} with cloud hole fractions of 0, 25, 50, 75, and 100\%.  The atmosphere models have effective temperatures 1000-2000K and use equilibrium chemistry with \fsed=2 and $\log g$=5.  \namesh A has properties consistent with a fully cloudy atmosphere.  \namesh B is most closely matched to a model with \teff=1150K and 25\% fractional hole coverage.
\label{fig_partly_cloudy}}
\end{figure}

\namesh B falls in the transition region between the late L dwarfs and the early T dwarfs, where models with partial cloud coverage  can be used to describe the source.  We use the models of \citet{2010ApJ...723L.117M} to infer an alternate physical description of this source.  These models, shown in Figure~\ref{fig_partly_cloudy}, compute the emitted flux for brown dwarfs with fixed masses and temperatures, but having different fractions of cloud coverage.  All models have \fsed=2, $\log g$=5.0, and equilibrium chemistry.  As expected, \namesh A is consistent with having a completely cloudy atmosphere (no holes in the cloud decks).  \namesh B has \teff $\approx$1200K and hole fraction=25\% for this set of models.   If a different set of base models was used, then the best fitting \teff\ and hole fraction would be different \citep{2010ApJ...723L.117M}, so caution should be used when implying a definite atmospheric structure based on this one case.  Nonetheless, it is important to point out that the \teff\ inferred by the partly cloudy models is consistent with the \teff\ inferred from the cloudy models with varying degrees of \fsed\ and atmospheric mixing.  

\subsubsection{Comparison to Other Brown Dwarfs}
\begin{figure}
\epsscale{0.6}
\plotone{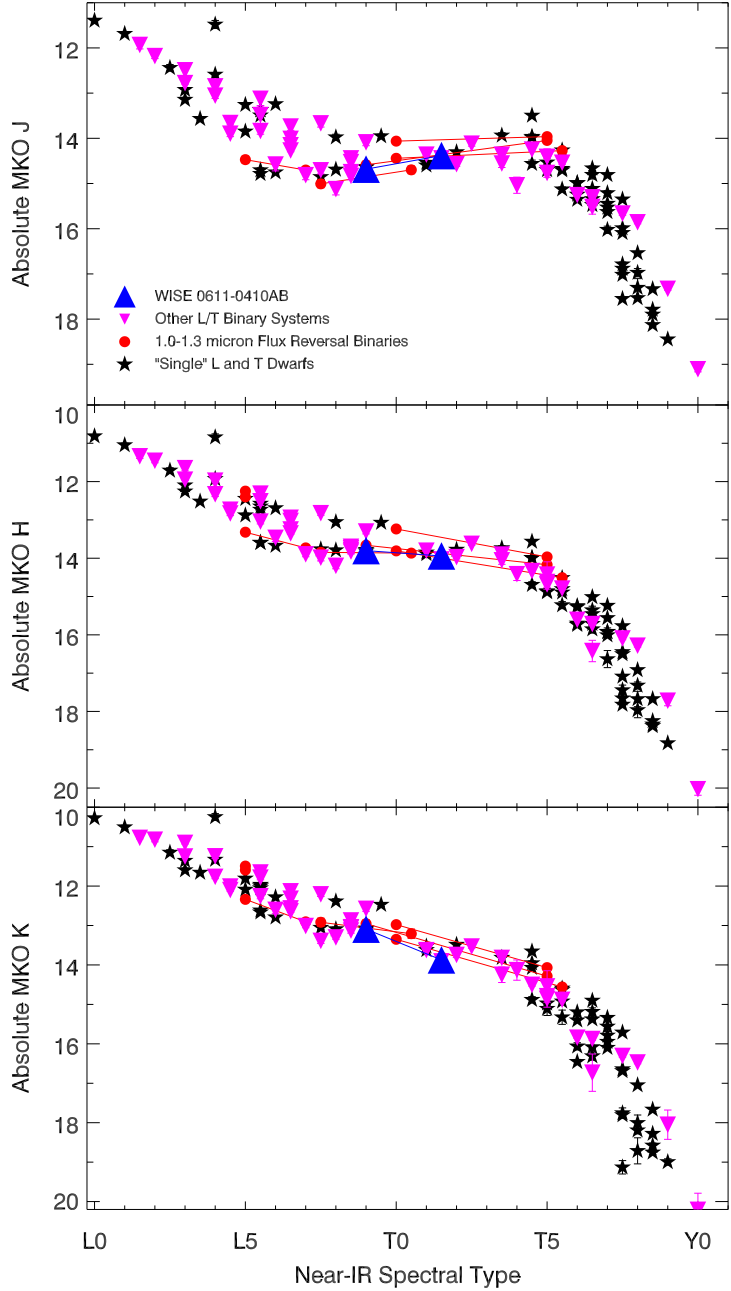}
\caption{Absolute MKO $J$ (top), $H$ (middle), and $K$ (bottom) magnitudes as a function of near-IR spectral type.  \namesh\ is denoted by the blue triangles.  Other 1.0--1.3 $\micron$ flux reversal binaries are shown with red circles; binary components are connected by a red line.  Components of other binaries are shown with magenta, inverted triangles.  Brown dwarfs not known to be binaries are given by black stars.  Data from the other L and T dwarfs are taken from the compilation of  \citet[][and references therein]{2012ApJS..201...19D} and \citet{2013arXiv1303.7283B}, from which young and metal poor objects have been excluded.  \label{fig_absjhk}}
\end{figure}

\begin{figure}
\epsscale{1.0}
\plotone{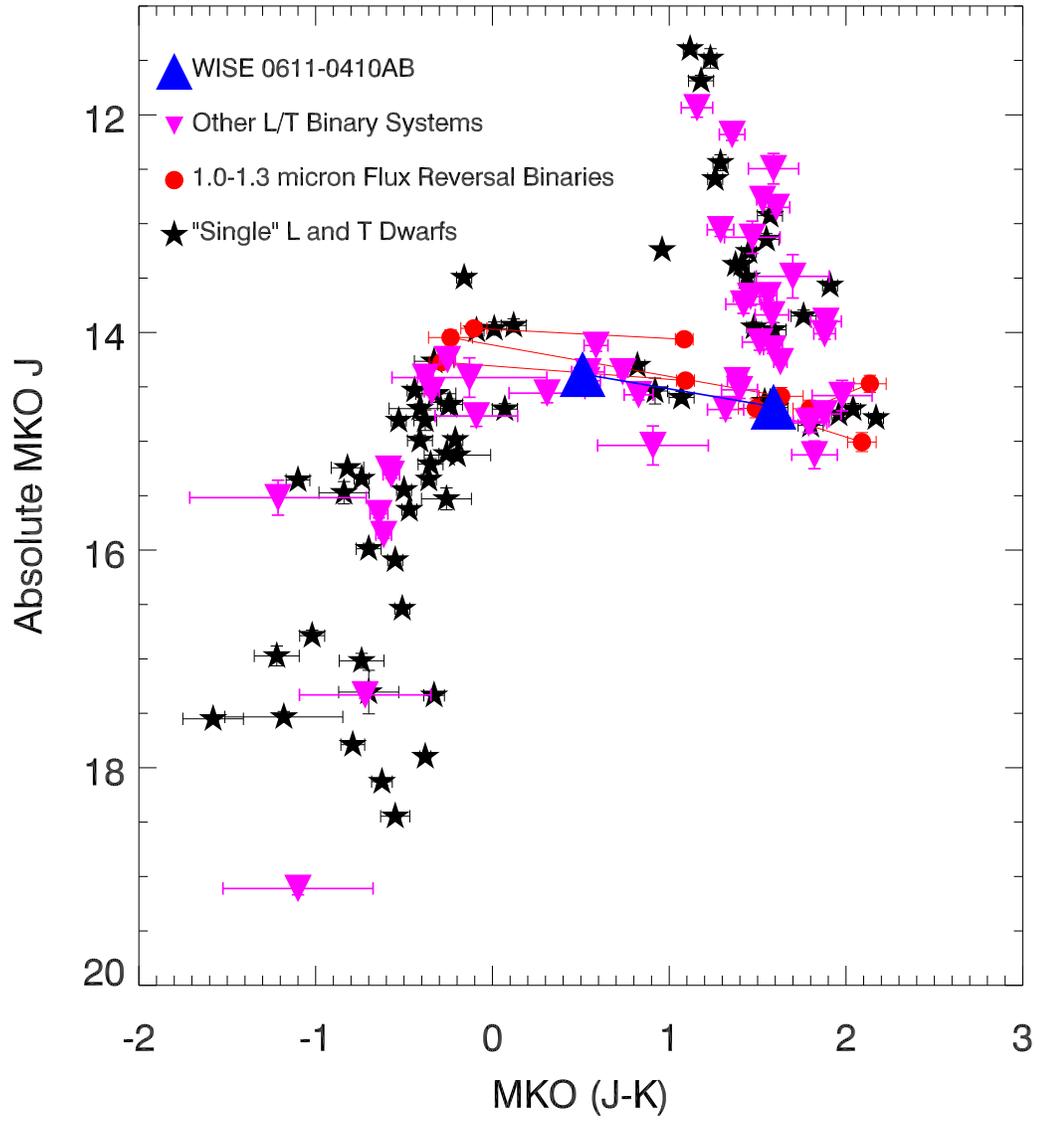}
\caption{Absolute MKO $J$ magnitude as a function of MKO $J-K$.  Data and point types are the same as in Figure~\ref{fig_absjhk}.\label{fig_absj_vs_jk}}
\end{figure}

The resolved photometry and parallax measured here allow us to compare \namesh AB to other brown dwarfs.  In Figure~\ref{fig_absjhk} we use the absolute magnitudes and spectral types estimated from the spectrum deconvolution to compare \namesh AB to other L and T dwarfs compiled by \citet[][and references therein]{2012ApJS..201...19D} and \citet{2013arXiv1303.7283B}.  Both components appear to be normal with respect to similarly-typed brown dwarfs.  However, since both spectral types are determined from the deconvolution of the composite spectrum, they should only be regarded as estimates.  The color-magnitude diagram shown in Figure~\ref{fig_absj_vs_jk} bypasses the spectral type uncertainty by making the comparison based entirely on the well-measured photometry and parallax.  The photometry of \namesh AB is well-matched to the photometry of objects with similar spectral types and colors, indicating that \namesh AB does not have significantly different metallicity or cloud properties compared to most other brown dwarfs.



\subsubsection{Orbit Estimations}
In order to estimate the orbital period of this system we first need to estimate the orbital semi-major axis and the masses of the components.  Since we have not detected any significant orbital motion of the secondary, we need to estimate the semi-major axis using a conversion factor from the projected separation to semi-major axis.  We use the conversion given in \citet{2011ApJ...733..122D} rather than the more commonly used conversion in \citep{1992ApJ...396..178F} because the former is based on very low-mass star and brown dwarf binaries, whereas the latter is based on higher mass, stellar binaries.   \citet{2011ApJ...733..122D} compute several conversion factors based on the discovery method and binary type.  For these observations, we use the ``moderate discovery bias,'' for which the inner working angle of our observations is approximately half of the true semi-major axis.  Using the factor for all very low-mass binaries ($a/\rho$=1.08) and the projected separation from the most recent epoch (8.1 AU),  we estimate the true semi-major axis to be 8.8$\pm$0.5 AU.  
The mass estimates are done using the model fitting described in the previous sections and are presented in Table~\ref{tab_best_fit_models}.  The orbital period for the best-fitting model parameters is then 78$\pm$8 years.

\section{Conclusions}
We have discovered a new brown dwarf binary system that straddles the L/T dwarf transition.  Resolved, differential photometry shows that the earlier-type primary is {\it fainter} than the cooler secondary in the $Y$ and $J$ bands, but brighter in $H$ and $K_s$, making this one of six known systems that shows a flux reversal in the 1.0-1.3 \micron\ region.   Although we do not have resolved spectroscopy of the components, deconvolving the composite spectrum into two spectra indicates the best-fit spectral types are L9 and T1.5.    We also present a new parallax of the system placing it at 21.2$\pm$1.3 pc.  Comparison to other L and T dwarfs shows that this system does not appear to have any unusual metallicity or age properties compared to the bulk of the field brown dwarfs, making it a good system to study for characterizing solivagant\footnote{Solivagant means ``wandering alone'' and is used to denote brown dwarfs that are isolated from other stars or brown dwarfs.} brown dwarfs.  However, with an orbital period estimated at 78 years, it will be some time before a good determination of the component masses can be made.

\acknowledgements
The authors acknowledge telescope operators Heather Hershley, Carolyn Jordan, Julie Renaud-Kim, and Cynthia Wilburn  
and instrument specialists Scott Dahm, Marc Kassis, and Luca Rizzi
for their assistance during the Keck observations.  We would also like to thank Chas Beichman for obtaining images of \namesh\ on 
2012 November 29 and 2013 September 21, Mark Marley and Didier Saumon for providing the models, Mike Read for help in mining the UKIDSS database and useful discussions, and the anonymous referee for useful comments.  Based partially on observations collected at the European Organisation for
Astronomical Research in the Southern Hemisphere, Chile program 186.C-0756.
RLS, HRAJ and FM would like to acknowledge the Marie Curie 7th European
Community Framework Programme grant n.247593  Interpretation and
Parameterization of Extremely Red COOL dwarfs (IPERCOOL) International
Research Staff Exchange Scheme. 
RAM acknowledges partial support from project PFB-06 CATA and from project IC120009 ``Millennium Institute of Astrophysics (MAS)'' of the Iniciativa Cientifica Milenio del Ministerio de Economa, Fomento y Turismo de Chile. 
CGT acknowledges the support of ARC grants DP0774000 and DP130102695.

This publication makes use of data products from 2MASS, which is a 
joint project of the University of Massachusetts and the Infrared Processing and Analysis Center/California Institute of 
Technology (Caltech), funded by the National Aeronautics and Space Administration (NASA) and the National Science Foundation.  This publication makes use of data products from the Wide-field Infrared Survey Explorer (WISE), which is a joint project of the University of California, Los Angeles, and the Jet Propulsion Laboratory (JPL)/Caltech, and NEOWISE, which is a project of JPL/Caltech.  WISE and NEOWISE are funded by NASA.
This work was supported by a NASA Keck PI Data Award, administered by the NASA 
Exoplanet Science Institute. Some of the data presented herein were obtained at the W. M. Keck Observatory from telescope time allocated to NASA 
through the agency's scientific partnership with the Caltech and the University of California. The Observatory was made possible by 
the generous financial support of the W. M. Keck Foundation. 
The authors recognize and acknowledge the 
very significant cultural role and reverence that 
the summit of Mauna Kea has always had within the 
indigenous Hawaiian community.  We are most fortunate 
to have the opportunity to conduct observations from this mountain.

{\it Facilities:} \facility{Keck:II~(NIRC2, LGS-AO)}, \facility{NTT~(SofI)}

\clearpage

\clearpage

\clearpage


\begin{deluxetable}{lccc}
\tablecaption{1.0-1.3 \micron\ Flux Reversal Binaries \label{tab_jflip}}
\tablewidth{0pt}
\tablehead{
  \colhead{Binary} &
  \colhead{SpT$_{\rm A}$} &
  \colhead{SpT$_{\rm B}$} &
  \colhead{Reference} \\
}
\startdata
\name{AB} & L9 & T1.5  & 1 \\
SDSS J102109.69$-$030420.1AB & T1 & T4 & 2 \\
WISE J104915.57$-$531906.1AB  & L7.5 & T0.5  & 3  \\
2MASS J14044941$-$3159329AB &  T1$\pm$1 & T5$\pm$1  & 4 \\
SDSS J153417.05+161546.1AB & T1.5 & T5.5  &5 \\
2MASS J17281150+3948593AB & L5 & L6.5 & 6, 7 \\
\enddata
\tablenotetext{\ }{References --- (1) this work; (2) \citet{2006ApJS..166..585B}; (3) \citet{2013arXiv1303.7283B}; (4) \citet{looper2008}; (5) \citet{liu2006}; (6) \citet{2003AJ....125.3302G}; (7) \citet{2010ApJ...710.1142B}}
\end{deluxetable}

\begin{deluxetable}{ccc}
\tablecaption{WISE~0611$-$0410 Photometry\label{tab_phot}}
\tablewidth{0pt}
\tablehead{
  \colhead{Property} &
  \colhead{Value} &
  \colhead{Reference} \\
}
\startdata
Spectral Type (near-IR) & T0 & 4 \\
2MASS $J$ & 15.489$\pm$0.055 & 1 \\
2MASS $H$ & 14.645$\pm$0.048 & 1 \\
2MASS $K_s$ & 14.221$\pm$0.070 & 1 \\
MKO $J$ & 15.398$\pm$0.006 & 3 \\
MKO $H$ & 14.743$\pm$0.005 & 3 \\
MKO $K$ & 14.292$\pm$0.005 & 3 \\
IRAC 3.6 $\mu$m  & 13.069$\pm$0.017 & 4 \\
IRAC 4.5 $\mu$m  & 12.924$\pm$0.017 & 4 \\
WISE $W1$\tablenotemark{a} & 13.559$\pm$0.026 & 5 \\
WISE $W2$\tablenotemark{a} &12.920$\pm$0.027 & 5\\
WISE $W3$\tablenotemark{a} & 11.926$\pm$0.282 & 5 \\
WISE $W4$\tablenotemark{a} & $>$8.619 & 5 \\
\enddata
\tablenotetext{a}{Its designation in the AllWISE Source Catalog is WISEA J061135.13-041024.1.}
\tablenotetext{\ }{References-- (1)2MASS; (2)this work; (3)UKIDSS; (4) \citet{2011ApJS..197...19K}; (5) AllWISE Source Catalog \citep{2013wise.rept....1C}}
\end{deluxetable}

\begin{deluxetable}{ccccc}
\tablecaption{NIRC2 Observing Log\label{tab_obs}}
\tablewidth{0pt}
\tablehead{
  \colhead{Date [UT]} & 
  \colhead{Filter\tablenotemark{a}} &
  \colhead{Exposure [sec]} &
  \colhead{FWHM [mas]} &
  \colhead{Strehl [\%]} \\
}
\startdata
2012 Jan 13 & $H$ & 1800 & 147 & 0.3  \\
2012 Apr 15 & $J$ & 180 & 75 & 2.5 \\
   & $H$ & 180 & 70 & 6.3  \\
   & $K_s$ & 180 & 67 & 14.4  \\
2012 Sep 06\tablenotemark{b} & $H$ & 30 & 100 & 2.0 \\
2012 Nov 29 & $H$ & 360 & 69 & 5.3 \\
2013 Sep 21 & $Y$ & 360 & 129 & 0.7 \\
    & $K$ & 360 & 98 & 9.2 \\
\enddata
\tablenotetext{a}{All filters are on the MKO system \citep{tokunaga2002}}
\tablenotetext{b}{Only a single image was obtained on this date.}
\end{deluxetable}

\begin{deluxetable}{cc}
\tablecaption{Parallax and Proper Motion of \namesh\label{tbl_plx}}
\tablewidth{0pt}
\tablehead{
  \colhead{Parameter}&
  \colhead{Value} \\
}
\startdata
$\alpha$, $\delta$ (J2000) & 6:11:35.0, $-$4:10:24.9 \\
Epoch [yr] & 2013.1629 \\
Absolute parallax [mas] & 47.25$\pm$3.22 \\
Distance [pc] & 21.2$\pm$1.3 \\
$\mu_{\alpha}$ [mas yr$^{-1}$] & 83.70$\pm$2.50 \\
$\mu_{\delta}$ [mas yr$^{-1}$] & $-$279.13$\pm$2.02 \\
Relative to absolute correction [mas] & 0.84 \\
Duration of observations [yrs] & 2.18 \\
reference stars, no. of observations &  291, 19 \\ 
\enddata
\end{deluxetable}

\begin{deluxetable}{lccc}
\tablewidth{0pt}
\tablecaption{Properties of \namesh{AB} \label{tab_binprop}}
\tablehead{
  \colhead{Parameter\tablenotemark{a}} &
  \colhead{A} &
  \colhead{B} &
  \colhead{$\Delta$}\\
}
\startdata
MKO $Y$ [mag] & \nodata & \nodata & $-$0.40$\pm$0.01 \\
MKO $J$ [mag]  & 16.31$\pm$0.01 & 16.01$\pm$0.01 & $-$0.30$\pm$0.01  \\
MKO $H$ [mag]  & 15.43$\pm$0.01  & 15.57$\pm$0.01 & +0.14$\pm$0.01 \\
MKO $K_s$ [mag] & \nodata & \nodata & +0.66$\pm$0.01 \\
MKO $K$ [mag]  & 14.72$\pm$0.01  & 15.50$\pm$0.01 & +0.78$\pm$0.01\\
MKO $J-K$ [mag] & 1.59$\pm$0.01 & 0.51$\pm$0.01 & \nodata \\
Est. Sp. Type                 & L9       &  T1.5 & \nodata \\
\enddata
\tablenotetext{a}{All magnitudes are apparent magnitudes.}
\end{deluxetable}

\begin{deluxetable}{ccccc}
\tablecaption{NIRC2 Astrometry\label{tab_sep}}
\tablewidth{0pt}
\tablehead{
 \colhead{} & \multicolumn{2}{c}{Measured} & \multicolumn{2}{c}{Predicted if BG object} \\
  \colhead{Date [UT]}  &
  \colhead{$\Delta$$\alpha$$\cos$($\delta$) [mas]} &
  \colhead{$\Delta$$\delta$ [mas]} &
  \colhead{$\Delta$$\alpha$$\cos$($\delta$) [mas]} &
  \colhead{$\Delta$$\delta$ [mas]} \\
}
\startdata
2012 Jan 13 & $-$177.2$\pm$15.8 & 346.8$\pm$5.8 & $-$177.2 & 346.8 \\
2012 Apr 15 & $-$184.4$\pm$6.5 & 353.4$\pm$3.2 & $-$205.7  & 424.3 \\
2012 Sep 06 & $-$180$\pm$10 & 349$\pm$10 & $-$234.3  & 530.1 \\ 
2012 Nov 29 & $-$182.5$\pm$3.7 & 346.6$\pm$4.9 & $-$256.1  & 591.9 \\
2013 Sep 21 & $-$181.8$\pm$3.4 & 338.6$\pm$1.9 & $-$322.6 & 808.6 \\
\enddata
\end{deluxetable}


\begin{deluxetable}{lcccc}
\tablecaption{Best Fit Models\label{tab_best_fit_models}}
\tablewidth{0pt}
\tablehead{
  \colhead{} &
  \colhead{\namesh A} &
  \colhead{\namesh B} 
}
\startdata
\chisq & 17.7$\pm$4.8 & 9.8$\pm$5.5  \\
mass [$M_{\rm Jup}$] & 64.8$\pm$0.4 & 52.1$\pm$12.0  \\
\teff\ [K] & 1300$\pm$26 & 1096$\pm$20  \\
\logg\ [cm s$^{-2}$] & 5.3$\pm$0.1 & 5.1$\pm$0.1 \\
\fsed & 2 & 4 \\
log \kzz & 0 & 6 \\
Approx. age [Gyr] & 1.5-4.0 & 1.0-2.0 \\
\enddata
\end{deluxetable}

\clearpage

\bibliographystyle{apj}
\bibliography{ms}

\end{document}